# Gradient Magnet Design for Simultaneous Detection of Electrons and Positrons in the Intermediate MeV Range

G. Tiwari,[1, §] R. Kupfer,[1] X. Jiao,[1] E. Gaul,[2] and B. M Hegelich[1, 3]

[1] *Center for High Energy Density Science, Department of Physics, University of Texas at Austin, Austin, 78712, USA*

[2]*National Energetics, 4616 West Howard Lane, Austin, 78728, USA*

[3]*Center for Relativistic Laser Science, Institute for Basic Science, Gwangju, 61005, South Korea*

We report the design and development of a compact electron and positron spectrometer based on tapered Neodymium Iron Boron magnets. We show that the tapered design forms a gradient magnetic field component allowing energy dependent focusing of the dispersed charged particles along a chosen detector plane using RADIA, a code developed by European Synchrotron Radiation Facility for solving three-dimensional magnetostatics configuration, and a fourth order Runge-Kutta Particle Tracking code. The mirror symmetric design allows for simultaneous detection of pairs i.e. electrons and positrons with energies from 2 MeV to 500 MeV. We have developed a prototype matching the design specifications. We investigate the effects of beam divergence on the energy resolution and signal conversion efficiency for a photo-stimulated luminescence-based Imaging Plates (IPs). The optimal entrance aperture of the magnet is found to be elliptical and bigger than that of conventional pinhole aperture-based spectrometer designs even for a divergent beam originating from a point source at 20 cm away (i.e. solid angle of ~8 milli steradians). The signal efficiency in BAS-IP of SR type ranges from 1% to 5% for a parallel beam incident on a circular aperture of 20 mm diameter type at a chosen detection plane whereas it drops by up to a factor of 3 in the presence of divergence of ~ 8 milli steradians. The proposed gradient magnet is suitable for the detection of low flux and/or monoenergetic type electron/positron signals with finite transverse sizes. It offers unparalleled advantages for gamma-ray spectroscopy in the intermediate MeV range.

**I. INTRODUCTION**

The introduction of chirped-pulsed amplification[1] in lasers has opened new avenues to explore fundamental physics in the uncharted area of electrodynamics ranging from generation of x-rays,[2, 3], acceleration of ions,[4, 5] neutrons,[6] and electrons[7,8] to creating astrophysical plasma like conditions in laboratories.[9] In a typical laser-matter experiment, the energy spectra of particles of interest are often characterized to unravel the physical phenomena involved. One of the simplest and widely used methods to measure the energy spectrum of the charged particles produced from a laser-matter interaction involves dispersion of these particles in a constant or uniform dipole magnetic field created either by an electromagnet or permanent magnets;[10–14] this technique relies on high flux of incident particles with specific angular and spectral distributions due to several restraining factors including, but not limited to, the use of a small pinhole to shield the detector plane from the background noise, relatively low response threshold of the detector being used and the size and shape of the compact detection system being implemented. For a low-repetition rate high intensity lasers such as the Texas Petawatt Facility,[15] these detectors tend to be less universal in terms of use and often need to be redesigned and reproduced to meet the demands of the variety of experiments being conducted at such a facility.[13] Although the digital adaption[11, 12] of the conventional designs make them appealing for high-repetition rate laser systems, it does not improve the efficiency for the detection of low flux electrons or positrons.

[§] Electronic mail: gtiwari30@physics.utexas.edu

Since multi-MeV photons have relatively small cross-sections in matter compared to the charged particles,[16] the issue of detection of low flux of multi-MeV photons has been overcome for Compton spectrometers in the past. Elegant magnet designs with the characteristic features of combined function magnets were implemented to properly characterize the energy spectrum of the incident photon beams as well as to maximize the acceptance of the electrons produced for better performance.[17–19] For example, Broad Range Electron Spectrometer (BRES)[18] implemented stigmatic focusing design based on a gradient field distribution that allowed both vertical and radial focusing of the dispersion of the Compton scattered electrons whereas Gamma to Electron Magnetic Spectrometer(GEPS)[19] adopted homogenous sectorial based design with electromagnets.[20,21] Both BRES and GEPS were designed to detect electrons only. Here we present a design based on tapered permanent magnets to improve the detection of low flux multi-MeV electrons and positrons produced from direct or indirect laser-matter experiments. The proposed design has the unique properties of a magnet desired for multi-MeV gamma spectroscopy.

The rest of the article is organized as follows: In section II, we present the design based on tapered permanent Neodymium Iron Boron (NdFeB) magnets and demonstrate its focusing properties. Section III proceeds with the development of a prototype featuring the simulated design based on the mapping of the magnetic flux density components. Section IV explores the issue and options of detecting charged particles along a curved plane and the dependence of the desired detection plane on the beam divergence. In section V, we present the analysis of acceptance, resolution and efficiency of the prototype based on the simulated design. We explore the extension of this design for the detection of multi-MeV photon beams in Section VI. Finally, we sum up our findings in section VI.

## II. UNIDIRECTIONAL FOCUSING FROM THE GRADIENT DESIGN

The tapered design shown in figures 1(a) and 1(b) was selected with a primary goal to design a compact magnetic spectrometer capable of detecting electrons and positrons with wide energy range from 2 MeV to 500 MeV. We used RADIA, a C++ object-oriented programming code interfaced to Mathematica and developed by European Synchrotron Radiation Facility, to solve the three-dimensional magnetostatics field configuration of the geometry.[22] RADIA uses boundary integral method and is computationally economic compared to commercial software based on the Finite Element packages.[22] The enclosure dimension of the final compact design is 250 mm by 320 mm by 136 mm; it consists of tapered NdFeB magnets as the magnetic poles which are represented by opaque red objects and the yoke made up of magnetic steel as indicated by the transparent grey surrounding the opaque poles as shown in Fig. 1. Each magnetic pole consists of two NdFeB blocks, each 220 mm long and 100 mm wide with 30 mm height for the first 25 mm and tapered down to a height of 10 mm over the remaining width of 75 mm at a constant angle of ~ 14.93 degrees. During the simulations, identical blocks with actual chamfers were applied to reproduce the realistic settings. The minimum gap between the magnet poles is 20 mm, for the first 25 mm from the center of the aperture as a zero reference, which increases linearly to the 60 mm till the pole edge. We note that the entrance aperture in the simulation was a square with the length of 20 mm whereas the actual prototype has a circular aperture with diameter 20 mm.

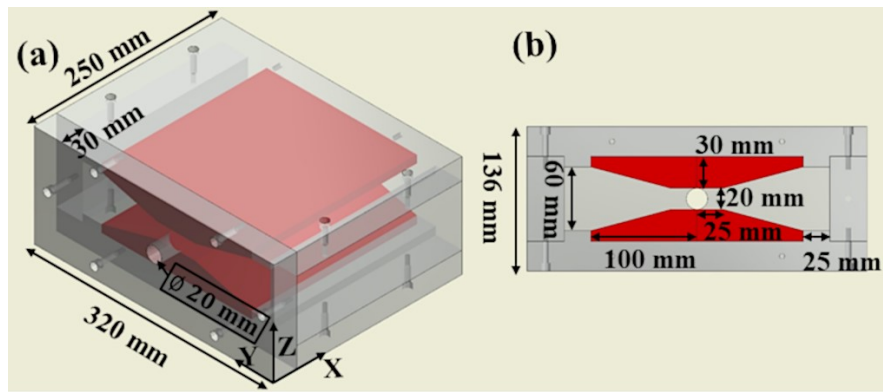

Fig. 1: Three-dimensional CAD Rendering of the Gradient Magnet shown in (a) Parametric View and (b) Back View.

Fig. 2(a) shows the surface plot of the vertical (z) component of the magnetic flux density ($B_z$) in the meridian plane (Z = 0) obtained from RADIA simulation for initially magnetized NdFeB blocks with residual magnetization ($B_r$) of 1.19 Tesla. The yoke initial magnetization was assumed to be zero to mimic the prototype. The maximum value of $B_z$ is 0.85 Tesla at X = 50 mm, Y = 0 and Z = 0. There is a dip in the $B_z$ profile at the center (Y = 0) followed by small peaks on either side along the X direction as observed in Fig. 2 (a). This is due to the presence of a small gap of ~0.3 mm between the magnetic poles that was considered to match with prototype's dimensional specifications. The profile has almost a constant gradient along Y on either positive or negative side and follows the tapering trend of the NdFeB blocks. Upon performing a linear fit to the $B_z$ profile from y = ±25 mm to ± 100 mm, we find the gradient field $\left(\frac{\partial B_z}{\partial y}\right)$ to be 6 T/m at X = -70 mm, 9.3 T/m at X = 0, 8.7 T/m at X =120 mm and 4.7 T/m at X = 140 mm. The $B_z$ profile shows the expected pattern of falling off over the front and back edges of the magnetic poles. The presence of the gradient component allows the energy dependent focusing of a spatially distributed beam over the entrance aperture upon entering the gap.[21] Figure 1 (d) shows the overlay of the electron and positron trajectories represented by the colored lines corresponding to discrete energies from 5 MeV to 500 MeV over the contour plot of the absolute magnetic flux density |B| in the meridian plane. The particle beams enter the magnet in the meridian plane at three different transverse locations i. e. X = -9 mm, 0 and 9 mm with zero divergence or parallel collimation. The sketch clearly indicates the focusing of individual energies at different positions in the gap of the magnetic assembly. The trajectories calculation was performed by solving the Lorentz force with the standard fourth order Runge-Kutta Particle Tracking code within a virtual box represented by the magnet gap for convenience.[23] Any interaction of the particles outside the virtual box boundary was ignored to let the particles propagate freely as shown in figure 2(b). The solid white line (for electrons) and dotted white line (for positrons) indicate the preferred detection planes (plane 1) for the incident beam; alternate detection planes (plane 2) represented by transparent black lines are also indicated for comparisons later.

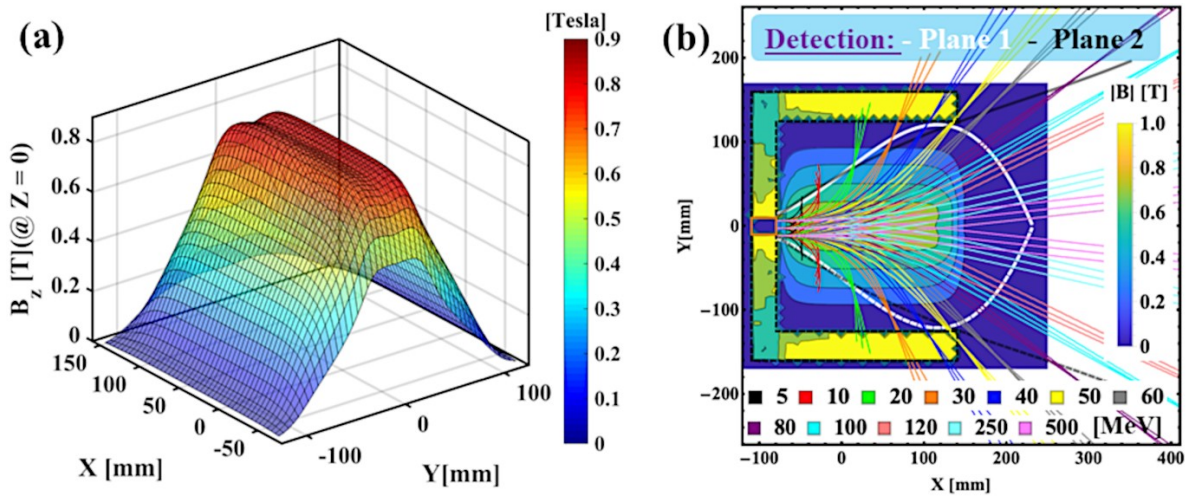

Fig. 2: (a) Surface Plot of the vertical component of the magnetic flux density in the meridian plane (Z =0). (b) Graphic display of particle trajectories with discrete energies from 5 to 500 MeV and Contour Plot of absolute flux density (|B|) in the meridian plane along with preferred (white lines) and alternate (transparent black) detection planes denoted as plane 1 and plane 2 respectively.

## III. NdFeB MAGNETIC ASSEMBLY AND FIELD MAPPING

The in-house assembled prototype along with the magnetic flux density measurement setup is shown in figure 2. The confining edges of the top and bottom yoke were made slightly larger to make assembling the NdFeB blocks in the yoke assembly easier. This resulted in the creation of a small gap of ~0.3 mm between the NdFeB blocks of same magnetic pole. In addition, four cylindrical Aluminum stubs each with diameter of 12.7 mm were introduced between the NdFeB blocks for structural integrity and magnetic stability. Two stubs are in the front side next to entrance aperture without blocking it and two are on the back as shown in figure 3. Since Aluminum is paramagnetic, the effect of the stub on the magnetic flux distribution are negligible.

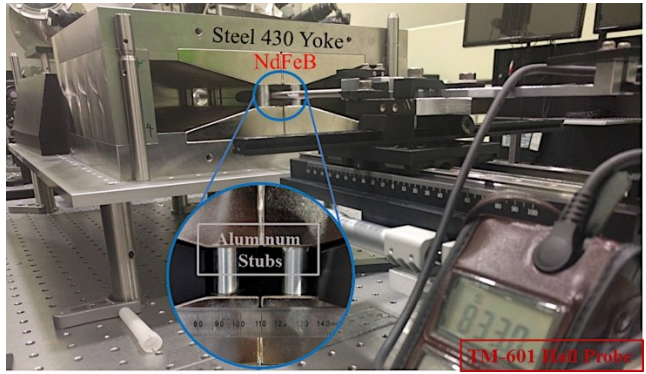

Fig. 3: Setup of the Gradient Magnet Prototype for magnetic flux density mapping.

We used a calibrated TM-601 Hall Probe to measure the vertical ($B_z$) and transverse horizontal ($B_y$) components of the magnetic flux density. The probe was setup in a three-dimensional translational stage comprising of a lab jack for vertical movements, two translation stages with rulings for longitudinal translation and a rail with rulings for horizontal transverse movements (see Fig. 3). Fig. 4(a) shows the plot of calculated (indicated by the lines) and measured (represented by "*o*") $B_z$ versus the longitudinal position (X) at different Y locations in the meridian plane. The fill around each line indicates the extent of the calculated $B_z$ due to the shift in Y by 3 mm. The maximum errors of the measured values are within 5% of the calculated values. The errors are prevalent on the tapered section with stronger flux density possibly due to the stronger curvature/tilt effect. In addition, slight asymmetry may have been introduced during the assembly process because of the fabrication and machining errors. Similarly, the measured values of the y-component of the magnetic flux density at X = 90 are found to agree with the calculated values within the errors bars introduced by vertical shift of 1 mm as shown in figure 4(b).

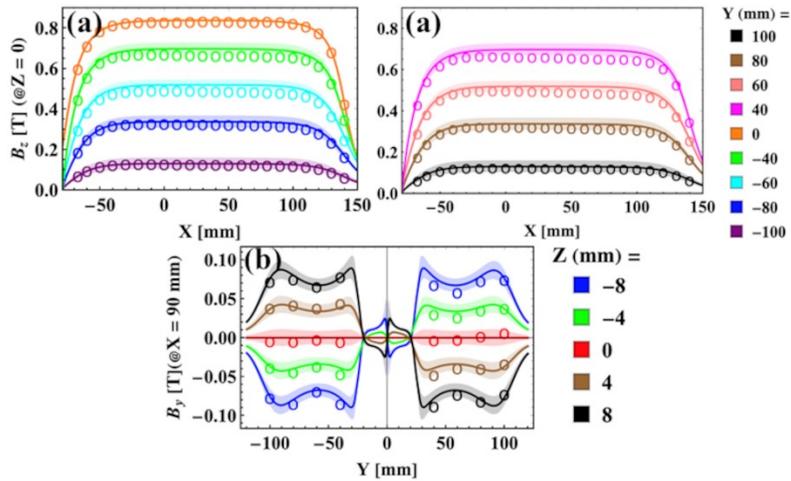

Fig. 4: Plots of calculated (indicated by the lines) and measured (represented by "o") (a) $B_z$ versus the longitudinal direction (X) at different Y locations and (b) $B_y$ versus the transverse horizontal direction (Y) at different heights.

## IV. DETECTION CHALLENGES AND OPTIONS

The tapered design offers strong focusing of electrons and positrons with energies up to 60 MeV within the geometrical bounds of the magnetic enclosure as clearly indicated by the presence of the gradient field on the order of few T/m (see Fig. 2(b)). However, the compact size of the design only allows weak focusing the particles above 60 MeV as shown in Fig. 2(b). In addition, the presence of incident beam divergence also results in weaker focusing strength of the magnet for all energies from 2 to 500 MeV. We demonstrate this effect by calculating the path length of a reference particle at the crossing point with other particles' trajectories incident on the magnet at a radius of 9 mm from the reference point (X= -80, Y= 0, Z=0). Since the focusing effect of the magnet is present only in the horizontal plane, we consider the particles distributed only in the horizontal plane at a radius of 9 mm from the reference point for convenience. Assuming the magnet to be a thin magnetic lens at the reference point, the crossing point of the trajectories for an energy would represent the focal length of the magnet for that energy. Figure 5(a) shows the plot of focal length versus energy of a particle beam with aperture radius of 9 mm for two cases of collimation, one with parallel collimation and the other with the divergence of 20 milli radians. The focal length of the particle beam increases drastically in the presence of beam divergence. There is a huge gap between the focal lengths of the particles with energies less than 60 MeV and greater than 70 MeV. The significant change in focal lengths in the presence of beam divergence makes a single shot-based data acquisition of the particle beam with continuous energy range from 2 to 500 MeV difficult.

Based on the focal lengths of a particle beam of various energies for no divergence, we identified two continuous detector planes as shown in Fig. 5 (b); plane 2 allows high resolution detection of particles up to 80 MeV whereas plane 1 covers the wide energy range with relatively limited or low resolution after 60 MeV. We chose plane 1 as the preferred detection plane to accommodate for the full energy range. The vertical extent of the detection plane is 20 mm as limited by the minimum gap between the magnet poles. Figure 5(b) shows the distribution of the electron hits in the detection plane 1 for various incident energies for an ideally collimated beam and for a beam originated from a point source at 20 mm away (with the solid angle of ~ 8 milli steradians) from the reference point of the magnet. The transverse spatial configuration of the incident beam at the entrance aperture of the magnet was kept fixed in both cases for consistent analysis.

A pixelated scintillator type adaption, as reported by *Gahn et. al*[11] and *Chen et. al.*[12], allows detection of the full energy range via discretization. Nevertheless, the response of the scintillator is energy dependent and the energy resolution cannot be maximized due to the fabrication limits on the smallest size of the scintillator. On the other hand, the Imaging Plates (IPs) are well calibrated over the wide range of energy spectrum compared to the scintillators;[24–26] they also offer high resolution and dynamic range, are unsensitive to electromagnetic pulses, durable and cost effective.[13,14,24,26] In addition, they are flexible and can easily move along a guided groove of plane 1 with no compromise in the detection of a particle beam with continuous energy range.

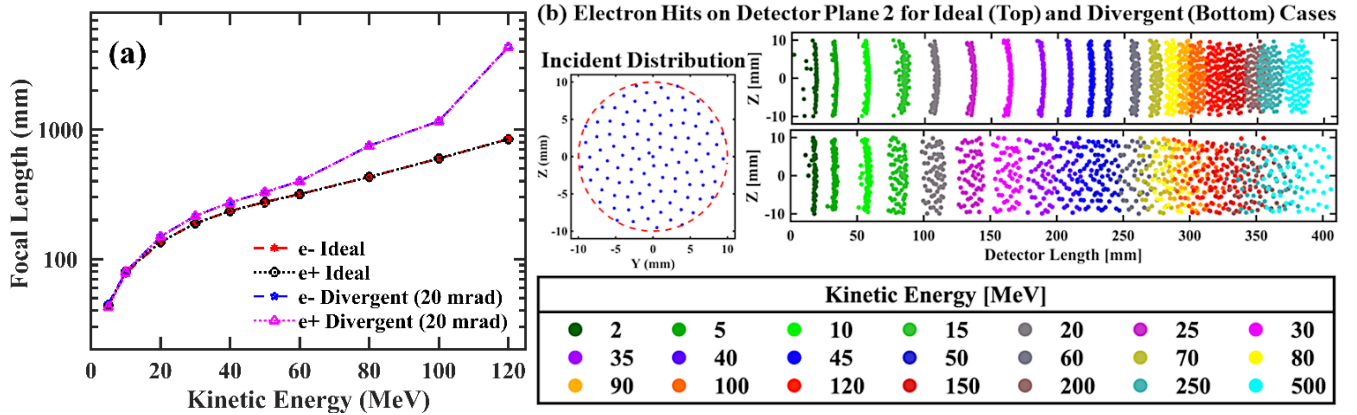

Fig. 5: (a) Plot of focal length versus energy of a horizontally distributed beam with an aperture radius of 9 mm for two cases of collimation, one with parallel collimation and the other with the divergence of 20 mrad. (b) The distribution of electrons hits in the detection plane 1 for various incident energies of an ideally collimated beam and of a beam originated from a point source at 20 mm away (has maximum divergence of ~ 8 msr) from the reference point of the magnet.

## V. RESOLUTION, ACCEPTANCE AND EFFICIENCY

Fig. 6 (a) shows the plot of energy of the particles versus the length of IPs with zero length starting at X =-80 mm for both ideal and divergent case of Fig. 5 (b). The energy follows the length of IP as an 8$^{th}$ order polynomial function along the detection plane 1 as given by $E(l) = 3.239\, l'^8 + 11.7\, l'^7 + 7.498\, l'^6 - 11.45\, l'^5 - 3.293\, l'^4 + 27.22\, l'^3 + 31.97\, l'^2 + 35.449\, l' + 43.31$. Here, $l' = (l - 218.3)/112.6$ and $l$ represents the length of IP. Similarly, the mean incident angles of the particles on the IP are found to be similar for both collimation cases as shown in figure 6 (b). The standard deviations of the incident angles are very high for particles with low energies up to 15 MeV because these particles experience strong magnetic force and sharp deflection via Lorentz force.

Even though the gradient magnet design offers nice feature of focusing effect for a horizontally spread beam, the acceptance of the gradient magnet is limited by the fact that the transverse spatial profile of a practical beam has both horizontal and vertical distribution. Based on the counting and tracking of one hundred particles incident on the magnet as shown in Fig. 5(b), we calculated the minimum Z-limit at the entrance aperture for all energies and collimation cases as shown in Fig. 6 (d). The minimal entrance aperture is found to be elliptical with major radius of 10 mm in the transverse-horizontal plane and a minor radius of 7. 25 mm for the parallel beam; for the divergent beam, the minor radius shrinks to 3.1 mm; this aperture is still bigger than that of typical pinhole 5 mm diameter aperture-based spectrometer designs.[11,14]

Now, we extend the charged particle tracking based calculations to calculate the expected signal efficiency in BAS- IPs of SR tpye[24,26] in the chosen detection plane 1. Fig. 6(d) shows the plot of expected photo-stimulated luminescence (PSL) signal per electron or positron versus the kinetic energy of the particles. The PSL sensitivity of the IP response for electrons are based on *Tanaka et. al.*[24] For the sake of simplicity, we assumed that the IP response to positrons were same as that to the electrons and did not take fading into account. The PSL efficiency in BAS-SR IPs ranges from 1% to 5% for a parallel beam incident on a circular aperture of 20 mm diameter with lower PSL signal evident in the energy range from 20 to 60 MeV (see Fig. 6(d)). We attribute the drop in the signal efficiency in this energy range to the smaller angle of incidence to the detection plane[24] (Fig. 5(b)) and reduction in the number of particles reaching the detector plane (see Fig. 5(c)). The likelihood of particle hits slowly rises with energy up to 15 MeV and starts to drop until it reaches a local minimum at 45 MeV before rising again to reach a plateau at energies greater than or equal to 200 MeV. The increase in hits at energies greater than 50 MeV is due to the vertical focusing introduced in the particle trajectories by the fringe fields as indicated by the particle hits in the detector plane 1 shown in Fig. 5(b). The trend is similar for beam incident with divergence of ~ 8 milli steradians on the the entrance aperture; however, the probability of hits is lower and PSL signal efficiency drops by a factor of 2 or higher at energy greater than 50 MeV compared to the parallel beam case. The PSL signal efficiency can be improved by about a factor of four by selecting BAS-IP of MS type over SR type based on the calibration data of various types of IP.[26]

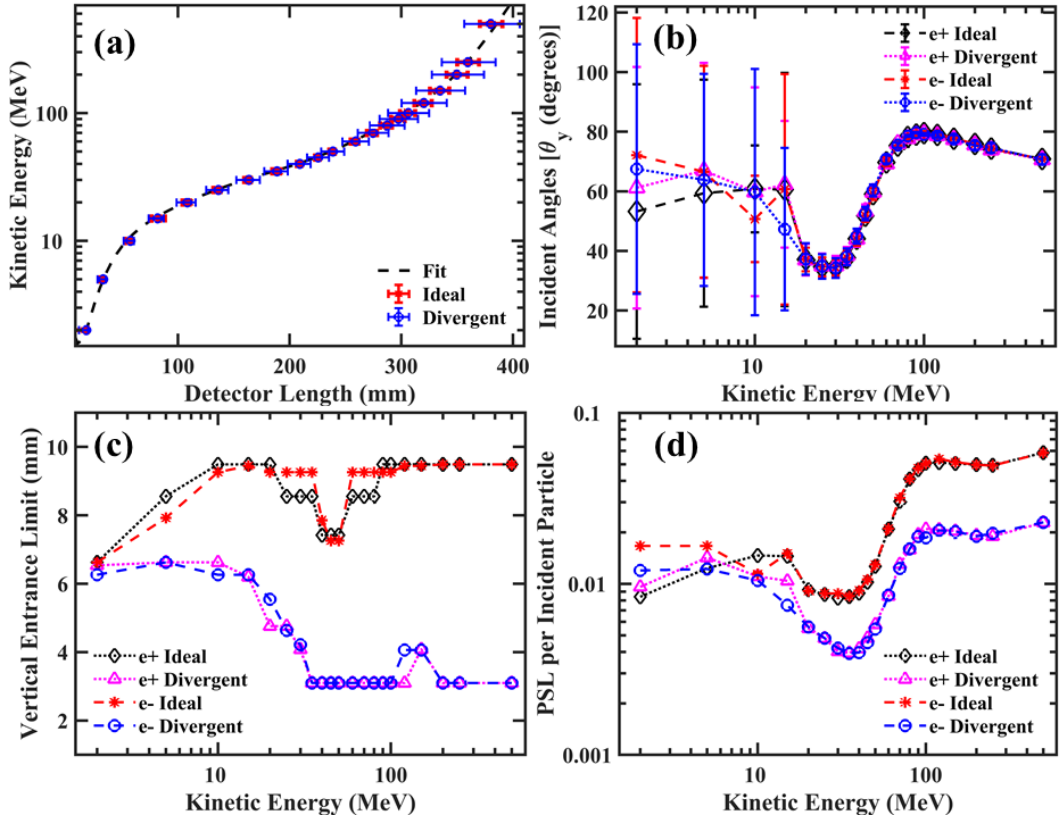

Fig. 6: Plot of (a) kinetic energy of the particles versus the length of the IPs with zero length starting at X =-80 mm for both ideal and divergent case of figure 5(b), (b) the mean incident angles of the particles on the IP versus their energies in MeV, (c) the minimum vertical limit at the entrance aperture versus the incident energy (MeV) for both collimation cases, and (d) PSL Signal Efficiency versus energies for based on the simulation of 100 particles incident on the entrance aperture of figure 5(b) and (d)

**VI. DISCUSSIONS AND CONCLUSIONS**

The gradient magnet has the unique properties of a magnet desired for multi-MeV gamma spectroscopy. First, it has the magnetic field configuration, like that of BRES[18] and GEPS[19] but that extends to higher energies, to allow for energy dependent focusing of an incident low flux electron/positron beam with finite transverse spatial profile. Second, its mirror symmetry allows detection of low flux positrons with energy resolution equal to that of electrons; this permits the removal of systematic error contained in the low-energy tail of the electron spectrum due to pair-production for accurate deconvolution of the energy spectrum of a low flux poly-chromatic gamma-ray beam with energy less than 20 MeV as discussed and proposed by *Corvan et. al*.[27] For instance, a pencil-like monoenergetic gamma-ray beam incident on a 20 mm Lithium block produces a peak electron signal with a conversion efficiency of 1.5% with kinetic energy of ~9 MeV whereas the peak positron signal occurs at around 5 MeV with a conversion efficiency of 0.08%.[27] On the other hand, the signal of electrons at 5 MeV is around 0.16% which is only two times as that of positrons. One alternative way to get rid of low energy tail (< 5 MeV) is to opt for thinner Lithium which is not favored when the incident photon beam has low luminosity.[27] For a lithium converter of 2 cm and $10^5$ photons of 10 MeV, a BAS-IP of SR type will register an integrated signal of about 0.08 PSL of positrons at 5 MeV in the detection plane 1 of the gradient magnet. For a MS type BAS-IP, the positron signal improves by a factor by 4 to ~3.2 PSL signal. Furthermore, since the pair-production dominates Compton scattering at energies greater than 20 MeV for medium Z materials like Aluminum or Titanium,[16] the complete reconstruction of the energy spectrum of a gamma-ray beam from pair-produced positrons at energies greater than 20 MeV is feasible. The gradient magnet offers extensive choice of converter materials and thicknesses to detect a poly-chromatic gamma ray beam depending on its energy range and luminosity.[28,29] We plan to submit a thorough article on the adaption of the gradient magnet as a gamma-ray spectrometer in near future.[28,29]

To sum up, we have demonstrated the design and development of a gradient magnet based on permanent magnets that can detect electrons and positrons with energies from 2 to 500 MeV. We have presented a thorough analysis based on trajectory calculations and practical particle beam considerations. We have shown that the performance and efficiency of the gradient magnet can be enhanced by using imaging plates and elliptical entrance apertures under tight collimation conditions. This magnet is suitable for detection of low flux and/or monoenergetic type electron/positron signals. These features make this magnet an appropriate candidate to be assembled into a gamma-ray spectrometer.[17–19,27]

**ACKNOWLEDGMENTS**

This work is performed under the auspices of the Air Force Office of Scientific Research FA9550-14-1-0045 and FA9550-17-1-0264. G. T. and R. K. acknowledge the support from the Glenn Focht Memorial Fellowship program. G. T. would like to thank Toma Toncian at the Helmholtz Zentrum Dresden-Rossendorf, Germany and Gilliss Dyer at the MEC Department of the LCLS facility, USA for insightful suggestions during the design and assembly process. The authors acknowledge unwavering support from the Machine Shop Staff at the Department of Physics of the University of Texas at Austin during the assembly of the prototype. G. T. also appreciates the support provided by the scientific and technical staff of the Center for Relativistic Laser Science, Institute of Basic Science, South Korea for the measurement of the magnetic flux density.